\documentclass{aa}

\usepackage[varg]{txfonts}
\usepackage{siunitx}
\DeclareSIUnit\um{\micro\meter}
\DeclareSIUnit\Msun{M_{$\odot$}}

\usepackage[tbtags,fleqn]{amsmath}
\usepackage{amsfonts}
\usepackage{tikz}
\usepackage{graphicx}
\usepackage{natbib}

\newcommand{\e}{\mathrm e}
\newcommand{\diff}{\mathrm d}

\newcommand{\R}{\mathbb R}
\newcommand{\mincir}{\raise
  -2.truept\hbox{\rlap{\hbox{$\sim$}}\raise5.truept \hbox{$<$}\ }}
\newcommand{\magcir}{\raise
  -2.truept\hbox{\rlap{\hbox{$\sim$}}\raise5.truept \hbox{$>$}\ }}


\DeclareMathOperator{\E}{E}
\DeclareMathOperator{\Hs}{H}

\DeclareMathOperator{\Var}{Var}
\DeclareMathOperator{\Cov}{Cov}

\bibpunct{(}{)}{;}{a}{}{,}

\begin{document}

\title{Fitting density models to observational data}
\subtitle{The local Schmidt law in molecular clouds} 
\titlerunning{Fitting density models to observational data -- The
  local Schmidt law}
\author{Marco Lombardi\inst{1} \and Charles J. Lada\inst{2} \and
  Jo\~ao Alves\inst{3}}
\authorrunning{M. Lombardi et al.}
\offprints{M. Lombardi}
\mail{marco.lombardi@unimi.it} 
\institute{%
  University of Milan, Department of Physics, via Celoria 16, I-20133
  Milan, Italy \and Harvard-Smithsonian Center for Astrophysics, Mail
  Stop 72, 60 Garden Street, Cambridge, MA 02138 \and University of
  Vienna, T\"urkenschanzstrasse 17, 1180 Vienna, Austria}
\date{Received ***date*** / Accepted ***date***}

\abstract{%
  We consider the general problem of fitting a parametric density
  model to discrete observations, taken to follow a non-homogeneous
  Poisson point process.  This class of models is very common, and can
  be used to describe many astrophysical processes, including the
  distribution of protostars in molecular clouds.  We give the
  expression for the likelihood of a given spatial density
  distribution of protostars and apply it to infer the most probable
  dependence of the protostellar surface density on the gas surface
  density.  Finally, we apply this general technique to model the
  distribution of protostars in the Orion molecular cloud and robustly
  derive the local star formation scaling (Schmidt) law for a
  molecular cloud.  We find that in this cloud the protostellar
  surface density, $\Sigma_\mathrm{YSO}$, is directly proportional to
  the square gas column density, here expressed as infrared
  extinction in the $K$-band, $A_K$: more precisely,
  $\Sigma_\mathrm{YSO} = (\num{1.65 \pm 0.19}) \,
  (A_K/\si{mag})^{\num{2.03 \pm 0.15}} \, \si{stars.pc^{-2}}$.}

\maketitle

\section{Introduction}
\label{sec:introduction}

In this paper we address a general statistical problem: the fit of
density models to discrete data.  Consider a non-negative function
$\rho(x | \theta)$, where $x \in \R^d$ is a point of a Euclidean space
of dimension $d$ (in typical situations $d = 1$, $2$, or $3$), and
$\theta$ is a vector of model parameters.  Suppose then that we
observe a realization of a non-homogeneous Poisson point process with
density\footnote{In this paper we decided to use a nomenclature closer
  to the one in use in the astronomical context.  For this reason, we
  will use the word ``density'' for the intensity of a Poisson point
  process, and we will use for this quantity the notation $\rho$
  instead of the standard statistical notation $\lambda$.  In this
  way, we hope to make the text clearer for the astronomical
  community.}  $\rho(x | \theta)$: that is, suppose that we observe a
random set of points $x_n \in \R^d$ distributed such that the number
of points $N_{A,B}$ in two disjoint sets $A$ and $B$ of $\R^d$ are
independent, and both $N_{A}$ and $N_B$ follow a Poisson distribution
$P(N) = \e^{-\mu} \mu^N/N!$ with means
\begin{align}
  \label{eq:1}
  \mu_A & {} = \int_A \rho(x | \theta) \, \diff^d x \; , &
  \mu_B & {} = \int_B \rho(x | \theta) \, \diff^d x \; .
\end{align}
We seek a way to infer the parameters $\theta$ from the points $\{ x_n
\}$.

The framework just introduced can be used to describe many different
situations.  For example, we could use it to model the stellar
distribution in globular clusters, using the King model for the
stellar positions; or we could use it to model the number counts of
galaxies.  \citet{1980ApJ...236...75S} considered the same framework
for fitting the distribution of galaxies in galaxy clusters.  In this
paper we use it to investigate the \textit{local} Schmidt law of star
formation in molecular clouds.  In all cases we would be interested in
inferring the relevant parameters of the density models used: center,
richness, and concentration parameter for the King model;
normalization, slope, and completeness limit for the number counts. In
general, the relevant parameters $\theta$ can be constrained using a
statistical frequentist approach, or a Bayesian approach.  In both
cases, we will need to evaluate the likelihood function, i.e.\ the
conditional probability to observe a given set of data points $\{ x_n
\}$ given the value of the parameters $\theta$.

\section{The log-likelihood}
\label{sec:log-likelihood}

Suppose that we observe a set of $N$ points $\{ x_n \}$ that we know
are a realization of a random Poisson point process with a given
density $\rho_1(x|\theta)$, which we assume to be normalized to
unity:
\begin{equation}
  \label{eq:2}
  \int \rho_1(x|\theta) \, \diff^d x = 1 \; ,
\end{equation}
Since this function, from a statistical point of view, is a
\textit{probability density}, we can immediately write the likelihood
of the configuration $\{ x_n \}$ given the parameters $\theta$ as
\citep{1980ApJ...236...75S}
\begin{equation}
  \label{eq:3}
  \mathcal{L}\bigl( \{x_n\} | \theta \bigr) = \prod_{n=1}^N
  \rho_1(x_n) \; .
\end{equation}
More generally, if we deal with a spatial density $\rho(x|\theta)$
that is \textit{not} normalized to unity, we can always consider the
function
\begin{align}
  \label{eq:4}
  \rho_1(x | \theta) & {} = \rho(x | \theta) / \mu \; , &
  \mu & {} \equiv \int \rho(x|\theta) \, \diff^d x \; ,
\end{align}
which by construction is normalized, and apply the
likelihood \eqref{eq:3} to $\rho_1$.  However, this procedure
prevents us from inferring any information on the normalizing constant
$\mu$; additionally, it has the drawback of forcing us to deal with
normalized densities, which in many contexts is not a natural choice.

To clarify these points, consider a simple model of star formation
in nearby molecular clouds, where we postulate that
\begin{equation}
  \label{eq:5}
  \Sigma_\mathrm{YSO} = \kappa \Sigma^\beta_\mathrm{gas}
\end{equation}
where $\Sigma_\mathrm{YSO}$ is the surface density of protostars
(stars pc$^{-2}$) in a molecular cloud and $\Sigma_\mathrm{gas}$ is
the mass surface density (\si{M_{$\odot$}.pc^{-2}}) of the molecular
gas.  This relation is similar to the version of the well-known
Schmidt Law \citep{1959ApJ...129..243S} introduced by
\citet{1998ApJ...498..541K} which relates the globally averaged
surface densities of the star formation rate ($\Sigma_\mathrm{SFR}$)
and the gas ($\Sigma_\mathrm{gas}$) in galaxies in a power-law
fashion.  Suppose that we know the locations $\{ x_n \}$ of $N$
protostars and the gas column density $\Sigma_\mathrm{gas}(x)$ in a
region of the sky, and that we intend to infer the parameters $\kappa$
and $\beta$.  In this case the likelihood discussed above does not
provide any information on $\kappa$, and this is unfortunate since
this parameter is directly linked to a critical piece of information,
the star formation rate of the molecular cloud.

In order to solve this problem, we note that the probability density
to observe $N$ points $\{ x_n \}$ from $\rho(x|\theta)$ can be
factorized as the probability to observe $N$ points (a Poisson
distribution with mean $\mu$), and the probability that these $N$
points be located at the positions $\{ x_n \}$ (as provided by
Eq.~\eqref{eq:3}):
\begin{equation}
  \label{eq:6}
  \mathcal{L}\bigl( [x_n] | \theta \bigr) = \e^{-\mu}
  \frac{\mu^N}{N!} \prod_{n=1}^N \rho_1\bigl(x_n|\theta\bigr) =
  \frac{1}{N!} \e^{-\int \rho(x) \, \diff^d x} \prod_{n=1}^N
  \rho\bigl(x_n|\theta\bigr) \; .
\end{equation}
In this expression the data points $[x_n]$ are taken to be
\textit{ordered}; if instead we consider the \textit{unordered set}
$\{ x_n \}$, we just have to multiply this result by the number of
permutations of $N$ elements, i.e.\ $N!$.  With this choice, the
log-likelihood takes the simple form
\begin{equation}
  \label{eq:7}
  \ln \mathcal{L}\bigl(\{x_n\} | \theta\bigr) = \sum_{n=1}^N \ln
  \rho(x_n | \theta) - \int \rho(x | \theta) \, \diff^d x \; .
\end{equation}

\section{Parameters inference}
\label{sec:parameters-inference}

Using Bayes' theorem
\begin{equation}
  \label{eq:8}
  P\bigl(\theta | \{ x_n \} \bigr) = \frac{\mathcal{L}\bigl( \{ x_n \} |
    \theta\bigr) p(\theta)}{\int \mathcal{L}\bigl( \{ x_n \} | \theta' \bigr)
    p(\theta') \, \diff \theta'} \; ,
\end{equation}
we can infer the posterior probability distribution of the parameters,
$P\bigl(\theta | \{ x_n \} \bigr)$ given the prior distribution
$p(\theta)$.  In our specific case, if we write $\rho(x|\theta) = \mu
\rho_1(x|\theta)$, then the likelihood factorizes into the product of
a term that depends only on $\mu$ and a term that depends on the other
density parameters $\theta$.  As a result, if the prior $p(\mu,
\theta) = p(\mu) p(\theta)$ also factorizes, then the posterior
distribution factorizes and $\mu$ is independent of $\theta$ (which
implies that the two are also uncorrelated).  In particular, if we use
the uninformative improper prior $p(\mu) = 1 / \mu$ for $\mu > 0$, we
find that the posterior for $\mu$ is a simple Gamma distribution
\begin{equation}
  \label{eq:9}
  P(\mu | \{ x_n \}) = \frac{\e^{-\mu} \mu^{N-1}}{\Gamma(N)} \; .
\end{equation}

Below, as often the case in astronomy, we will deal with an
unnormalized density $\rho$, and therefore we do not expect the
various parameters $\theta$ to be independent.

\section{Frequentist description}
\label{sec:freq-descr}

Alternatively, we can use to (log-)likelihood to obtain a
maximum-likelihood point estimate of the parameters $\theta$
\citep{1922RSPTA.222..309F}, i.e.\ find $\hat \theta =
\mathop{\mathrm{arg\ max}}_\theta \ln \mathcal{L}$.

Bounds on the errors on the parameters can be evaluated from the
Fisher information matrix (see, e.g., \citealp{ModStat}), that we
recall is defined as
\begin{equation}
  \label{eq:10}
  I_{ij} = \E \left[ \frac{\partial \ln \mathcal{L}}{\partial
      \theta_i} \frac{\partial \ln \mathcal{L}}{\partial
      \theta_j} \right] = - \E \left[ \frac{\partial^2 \ln
      \mathcal{L}}{\partial \theta_i \, \partial \theta_j} \right] \; ,
\end{equation}
where the symbol $\E$ denotes the operation of ensemble average.
The Fisher information matrix is related to the minimum
covariance matrix that can be attained by an unbiased estimator, as
provided by the Cram\'er-Rao bound:
\begin{equation}
  \label{eq:11}
  \Cov(\hat\theta) \ge I^{-1} \; .
\end{equation}
Since the maximum-likelihood estimator is asymptotically efficient
(i.e.\ it attains the Cram\'er-Rao bound when the sample size tends to
infinity) and the resulting errors on $\hat\theta$ tend to a
multivariate Gaussian distribution, it is interesting to obtain an
analytic result for the information matrix.  In Appendix~A.2 we show
that
\begin{align}
  \label{eq:12}
  I_{ij} & {} = \int \frac{1}{\rho(x|\theta)} \frac{\partial
    \rho(x|\theta)}{\partial \theta_i} 
  \frac{\partial \rho(x|\theta)}{\partial \theta_j} \, \diff^d x
  \notag\\
  & {} = \int \rho(x|\theta) \frac{\partial \ln
    \rho(x|\theta)}{\partial \theta_i} 
  \frac{\partial \ln \rho(x|\theta)}{\partial \theta_j} \, \diff^d x
  \; .
\end{align}

Finally, we can evaluate the goodness of the fit obtained by comparing
the maximum likelihood value with what we are expected to obtain.  The
expected (average) likelihood can be estimated analytically by
performing an ensemble average over the points $\{ x_n \}$.  As show
in Appendix~A.2, the average value of the log-likelihood calculated at
the \textit{true\/} value of the parameters is
\begin{equation}
  \label{eq:13}
  \E [ \ln \mathcal{L} ](\theta) = \int \rho(x | \theta)
  \bigl[ \ln \rho(x | \theta) - 1 \bigr] \, \diff^d x \; .
\end{equation}
For our purposes it is more relevant to obtain the expectation value
of the likelihood at its maximum value $\hat \theta$: this value can
then be compared with the observed maximum of the log-likelihood to
provide a goodness-of-fit proxy.  The result obtained is
\begin{equation}
  \label{eq:14}
  \E[\ln \mathcal{L}](\hat\theta) = \E[\ln \mathcal{L}](\theta) +
  \frac{J}{2} \; ,
\end{equation}
where $J$ is the number of the free parameters in the density, i.e.\
the dimensionality of the space of $\theta$.  

Finally, the variance of the log-likelihood is provided by
\begin{equation}
  \label{eq:15}
  \Var[ \ln \mathcal{L} ](\hat\theta) = \int \rho(x | \theta) \ln^2 \rho(x
  | \theta) \, \diff^d x \; .
\end{equation}
This expression, together with Eq.~\eqref{eq:14} is useful to define
acceptance boundaries for the likelihood value and therefore for model test.

\begin{figure*}
  \centering
  \includegraphics[height=0.33\hsize]{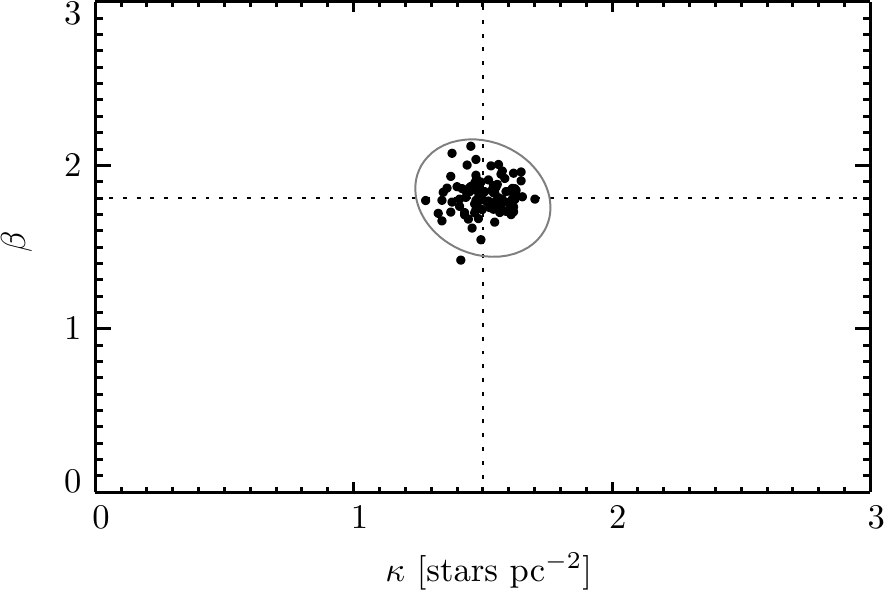}\hfill
  \includegraphics[height=0.33\hsize]{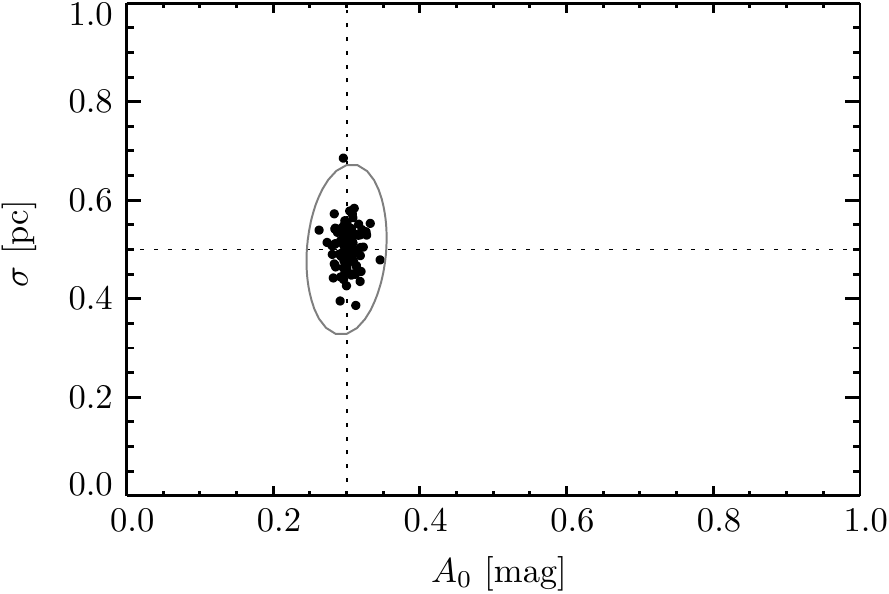}
  \caption{Maximum likelihood best-fit parameters for a set of 100
    simulations.  The dotted lines show the true, original values
    used: $\kappa = \SI{1.5}{star.pc^{-2}}$, $\beta = 1.8$, $A_0 =
    \SI{0.3}{mag}$, and $\sigma = \SI{0.5}{pc}$.  Note how the points,
    representing the individual best-fits, cluster around the true
    values.  The ellipses, showing the expected 3-$\sigma$ errors as
    deduced from the Fischer information matrix, are in excellent
    agreement with the observed distribution of points.} 
  \label{fig:1}
\end{figure*}

\section{Application: deriving the local Schmidt scaling law for a
  giant molecular cloud}
\label{sec:appl-deriv-local}

In this section we apply the framework developed above to one relevant
astrophysical problem, determining the local star formation scaling
law for a nearby Giant Molecular Cloud, which in its simplest version
can be written as in Eq.~\eqref{eq:5}.

In nearby molecular clouds $\Sigma_\mathrm{YSO}$ is a directly
measured quantity and can be related to $\Sigma_\mathrm{SFR}$ via
$\Sigma_\mathrm{SFR} = \Sigma_\mathrm{YSO} \times \langle m \rangle /
\tau$, where $\langle m \rangle$ is the mean mass of a protostar and
$\tau$ the age of the protostellar (class~I) population in the cloud.
We note here that we are interested in the \textit{local} star
formation scaling law within a GMC and this is physically different
from the Kennicutt-Schmidt law which relates \textit{global}
quantities averaged over entire galaxies (e.g.,
\citealp{1998ApJ...498..541K}) or subregions of galaxies (e.g.,
\citealp{2008AJ....136.2846B}) which in either case do not resolve the
individual molecular clouds.  Indeed, the indexes ($\beta$) of the
local and global relations are not necessarily physically related, nor
expected to be the same, when comparing individual molecular clouds
and normal galaxies (e.g., \citealp{2012ApJ...752...98C}).

Since we use infrared extinction measurements to trace the gas column
density in the clouds we write Eq.~\eqref{eq:5} as
$\Sigma_\mathrm{YSO} \propto A_K^\beta$ and note that
$\Sigma_\mathrm{gas} \propto A_K$ for a constant gas-to-dust ratio.
Additionally we allow for the possibility of a star-formation
threshold, i.e., a lower limit of the surface density for gas, or
equivalently dust, below which no star formation takes place and no
protostars are produced in situ (e.g., \citealp{2010ApJ...724..687L},
\citealp{2010ApJ...723.1019H}).  Finally, since it is physically
reasonable to assume that star formation is not an instantaneous
process, we allow for some diffusion of protostars from their birth
sites.  We model this diffusion process by smoothing the initial
protostellar surface density, $\Sigma_\mathrm{YSO}^{(0)}$, by a
Gaussian spatial kernel.  In summary we have
\begin{equation}
  \label{eq:16}
  \Sigma_\mathrm{YSO}(x|\kappa, \beta, A_0, \sigma) = \int \frac{1}{2
    \pi \sigma^2} \e^{|x - x'|^2 / 2 \sigma^2}
  \Sigma_\mathrm{YSO}^\mathrm{(0)}(x'|\kappa,\beta,A_0) \,
  \diff^2 x' \; ,
\end{equation}
where
\begin{equation}
  \label{eq:17}
  \Sigma_\mathrm{YSO}^\mathrm{(0)}(x|\kappa,\beta,A_0) = \kappa
  \Hs\bigl( A_K(x) - A_0 \bigr) \left( \frac{A_K}{\SI{1}{mag}}
  \right)^\beta(x) \; .
\end{equation}
In this equation $\Hs$ is the Heaviside step function (defined to be
zero for a negative argument and one for positive one).  The constants
involved are the normalization $\kappa$ (taken to be measured in units
of $\si{star.pc^{-2}.mag^{-\beta}}$, the star-formation threshold
$A_0$ (in units of $K$-band extinction), the dimensionless exponent
$\beta$, and the diffusion coefficient $\sigma$ (measured in pc).

The data at our disposal will be the extinction map of a molecular
cloud, i.e.\ $A_K(x)$, and a catalog with the positions of protostars
$\{ x_n \}$.  With these data we can fit the four parameters $\theta =
\{ \kappa, \beta, A_0, \sigma \}$ using both the approaches described
in Sects.~\ref{sec:parameters-inference} and \ref{sec:freq-descr}.

\subsection{Simulations}
\label{sec:simulations}

\begin{figure}
  \centering
  \includegraphics[width=\hsize]{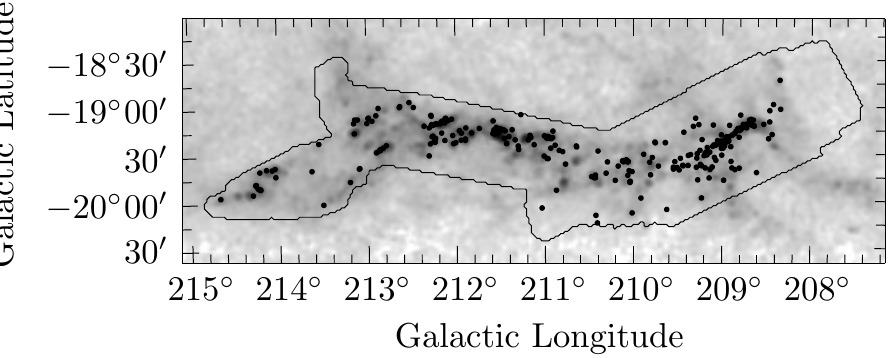}
  \caption{The extinction map of Orion~A (taken from
    \citealp{2011A&A...535A..16L}), together with the location of the
    Class~I protostars (from \citealp{2012AJ....144..192M}).  The
    polygonal line shows the region where the protostar observations
    have been carried out.}
  \label{fig:4}
\end{figure}

Before using the techniques described in this paper on real data, we
validated them on simulated data.  The simulations were carried out by
taking the extinction map of the Orion molecular cloud\footnote{We
  used from the original map only the area marked in Fig.~\ref{fig:4},
  corresponding to the region covered by the survey of protostars
  discussed below in Sect.~\ref{sec:application-orion}.} from
\citet{2011A&A...535A..16L}, and by randomly generating protostars
according to the law \eqref{eq:16}.  Specifically, we fixed the
parameters $\beta$, $A_0$, and $\sigma$ to test values, and we set
$\kappa$ such that the expected number of protostars in the field was
$300$.  For each simulation, we then drew the number of protostars
from a Poisson distribution with the appropriate average, and we
distributed the \textit{initial\/} positions of the stars in the field
following the density $\Sigma_\mathrm{YSO}^\mathrm{(0)}$ of
Eq.~\eqref{eq:17}, i.e.\ without any diffusion represented by the
$\sigma$ parameters.  Finally, we changed the initial positions of the
protostars by drawing random offsets from a two dimensional normal
distribution with variance $\sigma^2$, and by moving each protostar
position according to the drawn offsets.

We then tried to recover the ``unknown'' parameters using both a
maximum likelihood approach and a Bayesian inference.  The input data
provided to the code were the positions of the protostars, as
generated using the procedure described above, and the extinction map
of the cloud.  Figure~\ref{fig:1} reports the results obtained in a
typical set of simulation with true parameters $\kappa =
\SI{1.5}{star.pc^{-2}.mag^{-\beta}}$, $\beta = 1.8$, $A_0 =
\SI{0.3}{mag}$, and $\sigma = \SI{0.5}{pc}$.  To produce the plots,
we performed 100 independent simulations, drawing each time a
different set of protostars, and we report in the plot the locations
of the best-fit estimates obtained from the maximization of the
likelihood.  We also plot the expected 3-$\sigma$ error ellipse, as
derived from the Fisher information matrix.  As expected, and as shown
by these plots (and analogous ones produced during our tests), the
maximum likelihood estimate does not suffer any evident bias and is
able to constrain all parameters in a very efficient way.  Note that
for the set of simulations shown in Fig.~\ref{fig:1} we used on
average $300$ protostars, a number comparable with the number of
class~I objects known in Orion~A (see below).  During the tests we
also verified that the value of the likelihood function at its maximum
was compatible with the expectations of Eqs.~\eqref{eq:14} and
\eqref{eq:15}.

\begin{figure}
  \centering
  \includegraphics[width=\hsize]{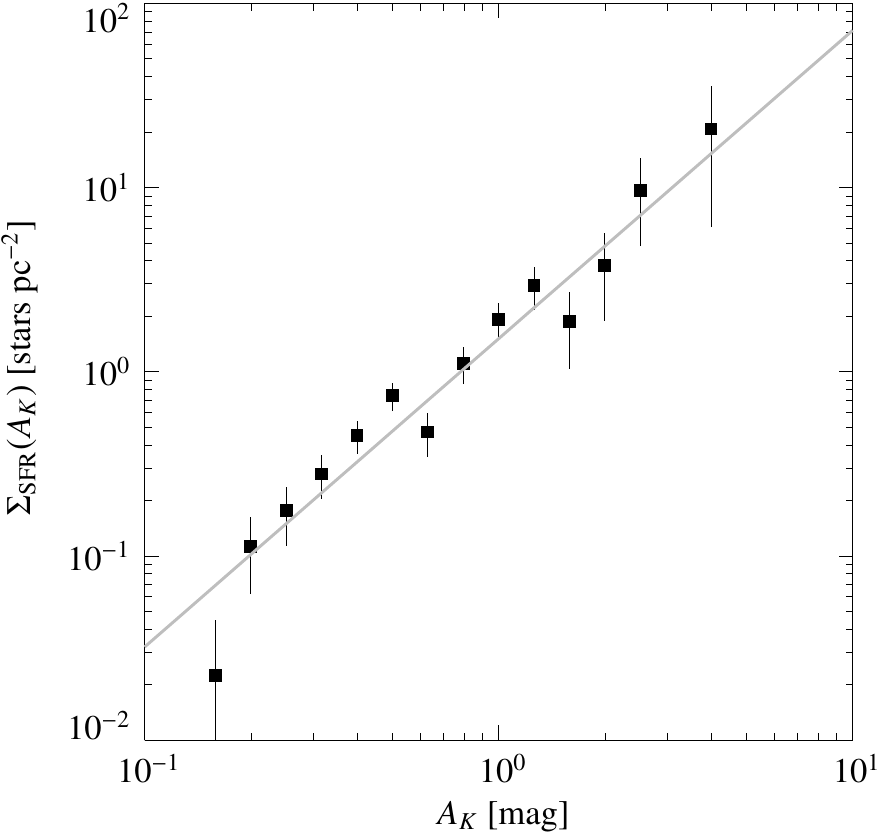}
  \caption{The differential distribution of $\Sigma_\mathrm{SFR}$ as a
    function of extinction on a simulation carried out with input
    parameters $\kappa = \SI{1.5}{star.pc^{-2}.mag^{-\beta}}$, $\beta
    = 1.8$, $A_0 = \SI{0.3}{mag}$, and $\sigma = \SI{0.5}{pc}$.  The
    best-fit power law, shown as a line in this plot, is $\kappa =
    \SI{1.23 \pm 0.12}{star.pc^{-2}.mag^{-\beta}}$ and
    $\beta = \num{2.71 \pm 0.08}$.}
  \label{fig:2}
\end{figure}

\begin{figure*}
  \centering
  \includegraphics[height=0.33\hsize]{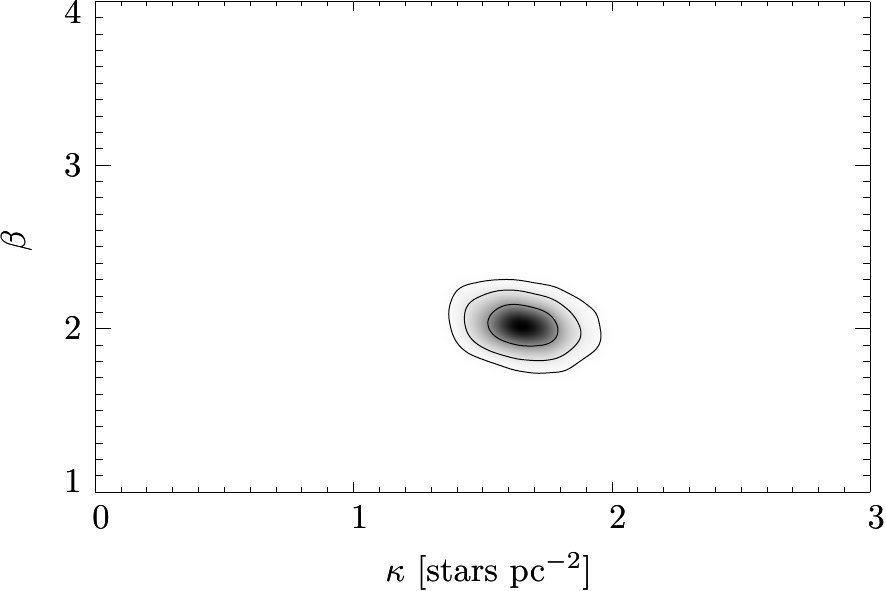}\hfill
  \includegraphics[height=0.33\hsize]{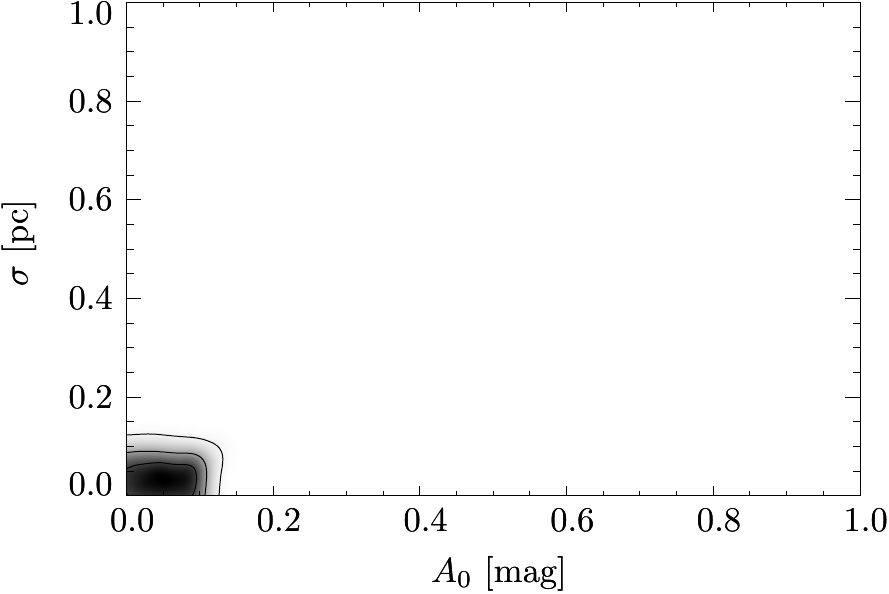}
  \caption{The posterior probability for the four density parameters
    obtained for the Orion~A molecular cloud.}
  \label{fig:3}
\end{figure*}

We also tested in a similar way the Bayesian approach.  For these
tests, we adopted both uniform and uninformative Jeffreys priors for
the parameters, excluding for all of them negative values.  In
general, during our tests we found that original, true values of the
parameters were always within the Bayesian $95\%$ credible intervals.

As a comparison, Fig.~\ref{fig:2} shows the projected density of
protostars as a function of the projected density of the dust.  To
make this figure, we considered increasing contour levels of
extinction, and we estimated the density of protostars within two
consecutive levels of extinction by computing the ratio between the
number of protostars and the area enclosed within the two contours;
errors were estimated from simple Poisson statistics.  The figure also
shows the best-fit power law obtained from these data, which is not in
agreement with the original input: the measured parameters where
$\kappa = \SI{1.23 \pm 0.12}{star.pc^{-2}.mag^{-\beta}}$ and $\beta =
\num{2.71 \pm 0.08}$, compared with the original input parameters of
$\kappa = \SI{1.5}{stars.pc^{-2}.mag^{-\beta}}$ and $\beta = 1.8$.
This simple test shows that in presence of a threshold and/or of a
diffusion of the protostars from their original loci of formation, the
standard fit of the $\Sigma_\mathrm{gas}$--$\Sigma_\mathrm{SFR}$ plot
leads to unreliable results.  Note that the poor performance of the
simple fit of histograms is a combination of several factors: the use
of a simple least-squares minimization algorithm does not take into
account the Poisson statistics of the bins (see
\citealp{nmeth0510-338}); the lack of modeling of the threshold and of
the diffusion; the lost of information when binning the data.  A
partial removal of these factors occasionally makes the algorithm less
biased (but only marginally so); in some cases (particularly when
adding the parameter $A_0$ to the fit) the bias was actually larger,
presumably as a result of a poor minimization of the least square
algorithm.  In general, we find that the fitting algorithm based on
binned data is very unstable, leading to large differences in the
results depending on the particular configuration used (i.e., on the
specific location of the protostars $\{ x_n \}$ and on the particular
choice of the bin sizes).  Note instead that the use of a Poisson
statistics with infinitesimal bins produces the same likelihood of
Eq.~\eqref{eq:6} (see Appendix~A.1), and therefore guarantees unbiased
and robust results.

Additionally, we also checked the results of the Bayesian analysis
when other observational effects were present in the data.
Specifically, after generating the stars according to the local
Schmidt law, we convolved the map with a Gaussian beam with
$\mathit{FWHM} = \SI{6}{arcmin}$ to simulate the effect of a
unresolved structures due to finite resolution.  Moreover, we added
artificial noise that takes into account the proper error propagation
in the extinction map (see \citealp{2001A&A...377.1023L}).
Simulations showed that the Bayesian technique can cope well with
these effects, and the results obtained are largely unaffected.

In summary, the simulations completely validated the approach
described in this paper.  We therefore applied our method to the best
studied star-forming molecular cloud, the Orion complex.

\subsection{Application to Orion~A}
\label{sec:application-orion}

In order to derive the local Schmidt scaling law in Orion-A, we used
our 2MASS/\textsc{Nicest} extinction map \citep{2011A&A...535A..16L}.
We preferred the \textsc{Nicest} \citep{2009A&A...493..735L} algorithm
over the \textsc{Nicer} \citep{2001A&A...377.1023L} one because the
former produces maps that are less affected by systematic biases in
the central regions of molecular clouds, where presumably most
protostars are formed.  Additionally, we used a catalog of protostars
obtained from a survey carried out in the region with Spitzer Space
Telescope \citep{2012AJ....144..192M}.  From this catalog we selected
only Class~I sources (specifically, objects classified as ``probable
protostars''); moreover, we excluded 56 sources found by
cross-correlation with a catalog of 624 foreground objects (identified
as unreddened sources visible in front of the molecular cloud, see
\citealp{2012A&A...547A..97A}).  As a result, we were left with 329
objects enclosed within the polygonal area in Orion-A where the
observations have been carried out (see \citealp{2012AJ....144..192M}
for details on the area selection).

We performed frequentist and a Bayesian analyses of these data using
the techniques described in this paper.  For the Bayesian analysis we
fitted the model \eqref{eq:16} with flat priors over all parameters
(taken however to be positive) using the likelihood of
Eq.~\eqref{eq:7}.  We explored the resulting posterior probability
distribution with Markov Chain Monte Carlo, using a simple
Metropolis-Hastings sampler.  The resulting posterior probability,
shown in Fig.~\ref{fig:3}, presents the following relevant results
(all bounds are referred to a $95\%$ confidence level interval):
\begin{itemize}
\item The exponent $\beta$ is exquisitely close to the value $2$: the
  result obtained is indeed $\beta = 2.03 \pm 0.15$.
\item We measure a star-formation coefficient $\kappa = \SI{1.65 \pm
    0.19}{star.pc^{-2}.mag^{-2}}$.
\item The data show that there is no threshold for star formation:
  $A_0 < \SI{0.09}{mag}$.
\item Class~I protostars seem to have undergone no detectable
  diffusion from their original positions beyond the resolution of our
  extinction maps.
\end{itemize}
The frequentist analysis provides consistent results.  Here we just
mention the fact that the best-fit log-likelihood is $\ln
\mathcal{L}(\hat \theta) = -1156$; as a reference, the value predicted
from Eq.~\eqref{eq:14} is $\E[\ln \mathcal{L}](\hat\theta) = -1137$,
with an expected tolerance computed from the square root of
Eq.~\eqref{eq:15} of $\sim 52$.  We deduce therefore that the model
proposed is in agreement with the observations.

\subsection{Discussion}

Using the analysis described above we can now write the fundamental
star formation scaling relation for the Orion A molecular cloud:
\begin{equation}
  \label{eq:18}
  \Sigma_\mathrm{YSO} = (\num{1.65 \pm 0.19}) \left(
    \frac{A_K}{\si{mag}} \right)^{\num{2.03 \pm 0.15}} \,
    \si{star.pc^{-2}} \; .
\end{equation}
This result can be considered a complete description of the local
Schmidt law characterizing this cloud because our analysis has enabled
the derivation of robust values for \textit{both} the power-law index,
$\beta$, and the coefficient, $\kappa$, the constant of
proportionality.  While $\beta$ determines how $\Sigma_\mathrm{YSO}$
varies with column density of the cloud, $\kappa$ sets the overall
scale or magnitude of the relation.

It is of interest to compare our results to those of a few recently
published studies.  \citet{2011ApJ...739...84G} derived a value for
$\beta$ of $1.8 \pm 0.01$ for the Orion cloud from a least-squares fit
to the $\Sigma_\mathrm{YSO}$--$\Sigma_\mathrm{gas}$ relation, using
the nearest neighbor method to derive $\Sigma_\mathrm{YSO}$ for each
star and infrared extinction measurements to derive the corresponding
$\Sigma_\mathrm{gas}$.  This result is in reasonable agreement with
our findings, particularly given the large scatter in the
\citet{2011ApJ...739...84G} data. Steeper $\beta$s ($\sim 4$),
however, have been derived for other clouds using different
methodologies \citep{2010ApJ...723.1019H, 2013ApJ...764..133H}.
Whether this represents evidence for variations in $\beta$ between
clouds is far from clear at this time.  More robust statistics and
application of a more consistent methodology to studies of other
clouds would be needed to make a better comparison to our results for
Orion and a better assessment of possible cloud-to-cloud variations in
the star formation scaling law.  We started already a follow-up
analysis in this direction, and the results of the application of the
statistical technique described in this paper to a set of clouds are
presented in \citet{2013arXiv1309.7055L}, to which we refer the reader
for a more the astrophysical conclusions of the analysis.

\subsection{Concluding Remarks}

In this paper we described the use of a likelihood analysis to derive
a parametric density model that best describes the discrete spatial
data.  The likelihood can be used both in a frequentist and Bayesian
analysis, and appears to be validated by its application to simulated
data in which input parameters were accurately recovered.  We applied
the method to model the observed distribution of Class I protostars in
the Orion A molecular cloud and derive the star formation scaling law
for that cloud. Specifically we find: $\Sigma_\mathrm{YSO} =
(\num{1.65 \pm 0.19}) \, (A_K/\si{mag})^{\num{2.03 \pm 0.15}} \,
  \si{stars.pc^{-2}}$.  Moreover, we found no evidence for an
  extinction threshold for star formation in the cloud and no evidence
  for any significant diffusion of the protostars from their birth
  sites.

\bibliographystyle{aa} 
\bibliography{../dark-refs}

\begin{thebibliography}{18}
\expandafter\ifx\csname natexlab\endcsname\relax\def\natexlab#1{#1}\fi

\bibitem[{{Alves} \& {Bouy}(2012)}]{2012A&A...547A..97A}
{Alves}, J. \& {Bouy}, H. 2012, \aap, 547, A97

\bibitem[{{Bigiel} {et~al.}(2008){Bigiel}, {Leroy}, {Walter}, {Brinks}, {de
  Blok}, {Madore}, \& {Thornley}}]{2008AJ....136.2846B}
{Bigiel}, F., {Leroy}, A., {Walter}, F., {et~al.} 2008, \aj, 136, 2846

\bibitem[{{Calzetti} {et~al.}(2012){Calzetti}, {Liu}, \&
  {Koda}}]{2012ApJ...752...98C}
{Calzetti}, D., {Liu}, G., \& {Koda}, J. 2012, \apj, 752, 98

\bibitem[{{Feigelson} \& {Babu}(2012)}]{ModStat}
{Feigelson}, E.~D. \& {Babu}, G.~J. 2012, Modern Statistical Methods for
  Astronomy (Cambridge University Press)

\bibitem[{{Fisher}(1922)}]{1922RSPTA.222..309F}
{Fisher}, R.~A. 1922, Royal Society of London Philosophical Transactions Series
  A, 222, 309

\bibitem[{{Gutermuth} {et~al.}(2011){Gutermuth}, {Pipher}, {Megeath}, {Myers},
  {Allen}, \& {Allen}}]{2011ApJ...739...84G}
{Gutermuth}, R.~A., {Pipher}, J.~L., {Megeath}, S.~T., {et~al.} 2011, \apj,
  739, 84

\bibitem[{{Harvey} {et~al.}(2013){Harvey}, {Fallscheer}, {Ginsburg}, {Terebey},
  {Andr{\'e}}, {Bourke}, {Di Francesco}, {K{\"o}nyves}, {Matthews}, \&
  {Peterson}}]{2013ApJ...764..133H}
{Harvey}, P.~M., {Fallscheer}, C., {Ginsburg}, A., {et~al.} 2013, \apj, 764,
  133

\bibitem[{{Heiderman} {et~al.}(2010){Heiderman}, {Evans}, {Allen}, {Huard}, \&
  {Heyer}}]{2010ApJ...723.1019H}
{Heiderman}, A., {Evans}, II, N.~J., {Allen}, L.~E., {Huard}, T., \& {Heyer},
  M. 2010, \apj, 723, 1019

\bibitem[{{Kennicutt}(1998)}]{1998ApJ...498..541K}
{Kennicutt}, Jr., R.~C. 1998, \apj, 498, 541

\bibitem[{{Lada} {et~al.}(2013){Lada}, {Lombardi}, {Roman-Zuniga}, {Forbrich},
  \& {Alves}}]{2013arXiv1309.7055L}
{Lada}, C., {Lombardi}, M., {Roman-Zuniga}, C., {Forbrich}, J., \& {Alves}, J.
  2013, ApJ, in press (astro-ph/1309.7055)

\bibitem[{{Lada} {et~al.}(2010){Lada}, {Lombardi}, \&
  {Alves}}]{2010ApJ...724..687L}
{Lada}, C.~J., {Lombardi}, M., \& {Alves}, J.~F. 2010, \apj, 724, 687

\bibitem[{{Laurence} \& {Chromy}(2010)}]{nmeth0510-338}
{Laurence}, T.~A. \& {Chromy}, B.~A. 2010, Nature Methods, 7, 338

\bibitem[{{Lombardi}(2009)}]{2009A&A...493..735L}
{Lombardi}, M. 2009, \aap, 493, 735

\bibitem[{{Lombardi} \& {Alves}(2001)}]{2001A&A...377.1023L}
{Lombardi}, M. \& {Alves}, J. 2001, \aap, 377, 1023

\bibitem[{{Lombardi} {et~al.}(2011){Lombardi}, {Alves}, \&
  {Lada}}]{2011A&A...535A..16L}
{Lombardi}, M., {Alves}, J., \& {Lada}, C.~J. 2011, \aap, 535, A16

\bibitem[{{Megeath} {et~al.}(2012){Megeath}, {Gutermuth}, {Muzerolle},
  {Kryukova}, {Flaherty}, {Hora}, {Allen}, {Hartmann}, {Myers}, {Pipher},
  {Stauffer}, {Young}, \& {Fazio}}]{2012AJ....144..192M}
{Megeath}, S.~T., {Gutermuth}, R., {Muzerolle}, J., {et~al.} 2012, \aj, 144,
  192

\bibitem[{{Sarazin}(1980)}]{1980ApJ...236...75S}
{Sarazin}, C.~L. 1980, \apj, 236, 75

\bibitem[{{Schmidt}(1959)}]{1959ApJ...129..243S}
{Schmidt}, M. 1959, \apj, 129, 243

\end{thebibliography}

\appendix

\section{Analytical derivations}
\label{sec:deriv-likel-prop}

\subsection{Alternative derivation of Eq.~\eqref{eq:7}}
\label{sec:altern-deriv-eq}

\begin{figure}
  \includegraphics[width=\hsize]{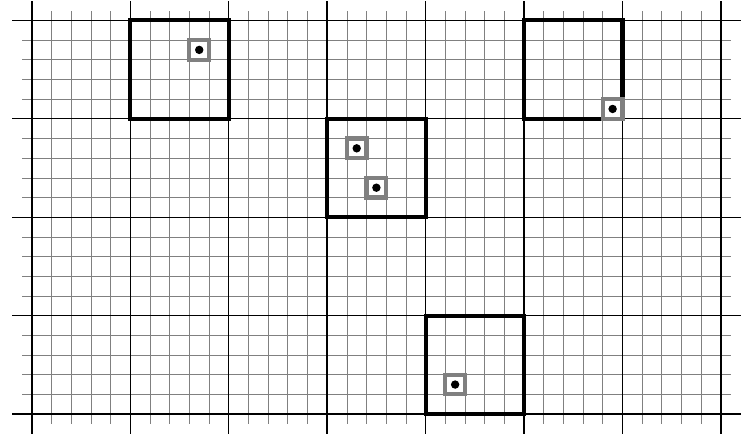}
  \caption{The derivation of the likelihood of Eq.~\eqref{eq:7}.  Note
    that when we reduce the grid size from the black grid to the grey
    one, each grid cell eventually contains at most one point.  Note
    also how the total area of the grid cells with points becomes
    negligible when the grid size is small.}
  \label{fig:5}
\end{figure}

In this section we provide an alternative derivation of
Eq.~\eqref{eq:7}, that highlights the relationship between the
likelihood and the fit of histograms.  Suppose we that we know
completely the density $\rho(x) \equiv \rho(x | \theta)$ responsible
for a Poisson point process.  One possibility to evaluate the
probability to observe a given configuration $\{ x_n \}$ of points is
to make a partition of the domain of $\rho(x)$ into a small regions or
bins of equal area $a$, chosen such that $a \ll 1 / \rho(x)$ for any
point $x \in \R^d$.  In this way, since the average number of stars
for each bin is much less then unity, then each bin will contain at
most one point.  In this limit, each bin will have with probability
$P_0(x) = 1 - a \rho(x)$ no stars, and with probability $P_1(x) = a
\rho(x)$ one star.  The joint probability for the whole configuration
will be therefore the product of the individual probabilities for the
individual bins, i.e.
\begin{equation}
  \label{eq:A.1}
  P\bigl(\{ x_n \} | \theta\bigr) = \prod_{n=1}^N \bigl[ a \rho(x_n |
  \theta) \big] \prod_m \bigl[ 1 - a \rho(x_m | \theta) \bigr] \; , 
\end{equation}
where the first product runs over the $N$ bins with a single data
point, and the second one over the other bins, i.e.\ all bins without
any data point (we recall that since bins are taken to be small, no
bin has more than a datapoint).  This probability is proportional to
$a^N$, and therefore as we take the limit $a \rightarrow 0$ it is
interesting to consider $\mathcal{L} = P / a^N$: that operation
corresponds to use a \textit{probability density}.  If we switch to
logarithms, the first product over $n$ becomes a sum of $\ln \rho$ at
the locations of the points, while the second product, in the limit of
$a \rightarrow 0$, becomes an integral over the entire space (because
the bins with points make a negligible contribution, see
Fig.~\ref{fig:5}):
\begin{equation}
  \label{eq:A.2}
  \ln \mathcal{L}\bigl(\{x_n\} | \theta\bigr) = \sum_{n=1}^N \ln
  \rho(x_n | \theta) - \int \rho(x | \theta) \, \diff^d x \; .
\end{equation}

\subsection{Properties of the log-likelihood}
\label{sec:prop-log-likel}

In order to prove Eq.~\eqref{eq:13}, we note that the average has to
be carried out over all possible configurations of data points $\{ x_n
\}$, and therefore involves only the first term, i.e. the summation,
of this equation; the integral is a constant term and the average is
trivially itself. Let us call $\ell$ the first term of
Eq.~\eqref{eq:13}, and $\mu$ the second, so that $\ln \mathcal{L} =
\ell + \mu$.  By using the same reasoning of Eq.~\eqref{eq:6}, we
can write the average of $\ell$ by considering separately the
probability to have exactly $N$ points in the field, which follows a
Poisson probability, and the probability for the distribution of each
of these points. In summary,
\begin{align}
  \label{eq:A.3}
  \E[\ell] & {} = \sum_{N=0}^\infty \e^{-\mu}
  \frac{\mu^N}{N!} \int \diff x_1 \, \frac{\rho(x_1)}{\mu}
  \cdots \int \diff x_N \, \frac{\rho(x)}{\mu} \ell \;
  \notag\\
  & {} = \sum_{N=0}^\infty \frac{\e^{-\mu}}{N!} \int \diff x_1 \,
  \rho(x_1) \cdots \int \diff x_N \, \rho(x_N) \sum_{n=1}^N \ln
  \rho(x_n) \notag\\
  & {} = \sum_{N=0}^\infty \e^{-\mu} \frac{\mu^{N-1}}{N!}
  \sum_{n=1}^N \int \diff x_n \, \rho(x_n) \ln \rho(x_n) \notag\\
  & {} = \sum_{N=1}^\infty \e^{-\mu}
  \frac{\mu^{N-1}}{(N-1)!} \int \diff x \, \rho(x) \ln \rho(x)
  \notag\\
  & {} = \int \diff x \, \rho(x) \ln \rho(x) \; .
\end{align}
Finally, putting back the term $\mu$ of $\ln \mathcal{L}$ we obtain
the desired result of Eq.~\eqref{eq:13}.

The variance of $\ln \mathcal{L}$ is identical to the variance of
$\ell$, which can be evaluated from $\E[\ell^2]$.  Following a
calculation similar to Eq.~\eqref{eq:3} and using the independence of
$x_n$ from $x_m$ for $n \ne m$ one obtains
\begin{align}
  \label{eq:A.4}
  \E[\ell^2] = \left[ \int \rho(x) \ln \rho(x) \, \diff^d x \right]^2 +
  \int \rho(x) \ln^2 \rho(x) \, \diff^d x \; .
\end{align}
From this equation one immediately derives $\Var(\mathcal{L}) =
\Var(\ell) = \E(\ell^2) - \E^2(\ell)$ as given in Eq.~\eqref{eq:15}.

In a similar way one can derive Eq.~\eqref{eq:12}.  We start from the
second equality of Eq.~\eqref{eq:10}, and we note that the partial
derivatives of $\ln \mathcal{L}$ can be written as
\begin{align}
  \label{eq:A.5}
  -\frac{\partial^2 \ln \mathcal{L}}{\partial \theta_i \, \partial
    \theta_j} = {} & \sum_{n=1}^N \left[ \frac{1}{\rho^2(x_n)}
    \frac{\partial \rho(x_n)}{\partial \theta_i} \frac{\partial
      \rho(x_n)}{\partial \theta_j} - \frac{1}{\rho(x_n)}
    \frac{\partial^2
      \rho(x_n)}{\partial \theta_i \, \partial \theta_j} \right]
  \notag\\
  & {} - \int \frac{\partial^2 \rho(x_n)}{\partial \theta_i
    \, \partial \theta_j} \, \diff^d x \; .
\end{align}
When taking the average over the point positions $\{ x_n \}$ of this
expression, the summation of the first line becomes an integral over
$\rho(x) \, \diff^d x$, and as a result the last term of the first
line cancels with the integral of the second line.  We are therefore
left out with Eq.~\eqref{eq:12}.

%

\end{document}